High Quality Factor Silicon Nitride Nanomechanical Resonators Fabricated by Maskless Femtosecond Laser Micro-machining

Roghayeh Nikbakht[1,2,*], Xitong Xie [2], Arnaud Weck[2,3,4], Raphael St-Gelais[2,3,4]

[1] Department of Mining and Materials Engineering, McGill University, 3610 Rue University, Montréal, QC, H3A 0C5, Canada

[2] Department of Mechanical Engineering, University of Ottawa, 161 Louis Pasteur, Ottawa, ON, K1N 6N5, Canada

[3] Department of Physics, University of Ottawa, 150 Louis Pasteur, Ottawa, ON, K1N 6N5, Canada

[4] Centre for Research in Photonics at the University of Ottawa, 25 Templeton St., Ottawa, ON, K1N 6N5, Canada

* Corresponding author, Roghayeh.Nikbakht@McGill.Ca

**Abstract**

Freestanding Silicon nitride (SiN) devices are central to the field of nanomechanical resonators and for other technology applications such as transmission electron imaging and nanopore bioassays. The nanofabrication techniques used for fabricating these devices often lack flexibility. While photolithography requires printing of an expensive photomask for each new design iteration, electron-beam lithography is extremely slow and commands high equipment cost. Here we demonstrate maskless rapid prototyping of freestanding SiN nanomechanical resonators fabricated by femtosecond laser ablation of plain SiN membrane in ambient air. We fabricate microbeams with different widths from 7 to 100 µm, and we characterize their resonance frequency and mechanical quality (Q) factors. We find that membrane cracking can be avoided during fabrication by carefully engineering the etch pattern, and that laser etching has a negligible effect on built-in tensile stress. For each beam, Q-factors are measured for several eigenmodes and are found to remain high after laser etching. All beams show quality factors greater than $10^5$, while unetched plain membranes have $Q > 10^6$. Possible causes for Q-factor reduction are identified, along with future process improvement directions.

**Main text**

Low-pressure chemical vapour deposited (LPCVD) silicon nitride (SiN)[1,2], is a material of platform of central importance in a large variety of high technology applications. In nanomechanical resonators—i.e., the topic of the current work—the high tensile stress of SiN greatly dilutes material damping, leading to record-high mechanical quality factor (Q-factor)[3] and operation in quantum coherent regimes[4–6]. In other technological fields, such as integrated optics, SiN thin films have been used to support extremely low loss guided modes, enabling applications[7] such as optical frequency combs and entangled photon pairs generation for quantum information processing [8]. SiN cantilevers have also been commercially available for biosensing for many years[9].

Traditional photolithography and e-beam lithography are used for fabricating virtually all SiN-based devices, even though these techniques often impose significant limitations and trade-offs in term of flexibility. On one hand, e-beam lithography allows great flexibility in varying devices geometries between each fabrication runs, but it is essentially limited to some specialized academic research facilities due to its extremely low speed and high equipment cost. On the other hand, traditional photolithography allows for very high throughput, but it is based on costly photomasks that cannot be modified or adjusted once produced by specialized vendors. Traditional photolithography is therefore well suited to large volume production, but is often prohibitively expensive and time-consuming for prototyping and small-volume production. In between these two extremes, a rapid, flexible, and low-cost method of direct patterning SiN devices does not exist, even though it could provide a solution that would benefit researchers and small businesses in a similar way as 3D printing did for other fields in recent years.

In this work, we investigate direct laser patterning (i.e., without lithography) of nanomechanical resonators in plain, freestanding SiN membranes—a starting platform that is widely available from multiple vendors[10,11]. Laser-based patterning is advantageous for its high ablation speed and its compatibility with atmospheric pressure operation[12–14]. Femtosecond lasers, with a pulse width shorter than the electron-phonon coupling time (1-100 ps, depending on the electron-phonon coupling strength of the material[15]), eliminate thermal diffusion in the vicinity of the laser exposed region and thereby prevent the creation of heat-affected zone[15]. A few reports exist on the use of short-pulsed lasers for ablating free-standing dielectric structures, but not for creating SiN nanomechanical resonators. Uesugia et al.[12,16] created holographic patterns and drilled nanoholes on a 10 nm thick SiN free-standing membrane using a femtosecond laser ($\lambda$= 1040 nm and pulse duration of 311 fs). Drilling nanoholes with one-tenth of the laser wavelength (100 nm), they showed that exploiting the beam intensity distribution and the material ablation threshold can allow features below the light diffraction limit. Shi et al.[17] used ultrafast lasers in the manufacturing of mechanical micro glass blown $SiO_2$. Femtosecond lasers are also used for ablation unsuspended SiN, such as passivation layers on textured silicon wafers used in photovoltaic cells[18,19]. Micro-machining of nanomechanical resonators on suspended membranes has yet to be reported and is the goal of the current study.

The silicon nitride nanomechanical resonators machined in this work are fabricated by laser ablation of free-standing low-stress LPCVD SiN square membranes of 100 nm nominal thickness (90 nm actual thickness was measured by ellipsometry[20]), and 1.2 × 1.2 mm side length. The membranes are fabricated in-house (see an 11-membranes chip in Fig. 1 b), but similar membranes can also be purchased commercially from various vendors[10,11]. Details of our membrane fabrication process can be found in [21].

A Light Conversion PHAROS laser outputting 513 nm light at a pulse duration of 320 fs and at a 200 kHz repetition rate is used for micromachining of SiN membranes (See Fig. 1a). The laser is focused onto the sample using a microscope objective (20× infinity corrected high working distance NIR Mitutoyo microscope objective with numerical aperture NA=0.4) down to a spot size of approximately 1 µm. All samples are machined as close as possible to the focal point to have the highest machining precision. The laser fluence on the membrane is adjusted during machining using a succession of two polarization beam splitters and is measured with an OPHIR VEGA digital power and thermal sensor (OPHIR 3A-P-Quad). The threshold of laser fluence for drilling holes through the SiN membranes is measured between 0.73-1.2 J/cm$^2$ per pulse (see Fig 2 b). The laser fluence is then set at a low multiple of this value (2.6 J/cm$^2$) to ensure each pulse completely drills through the SiN membrane in a repeatable fashion, while not applying unnecessary high optical power. This fluence value corresponds to an average incident power of 9.1 mW, or 46 nanojoules per pulse. The diameter of the drilled holes varies from 600 nm to 930 nm using laser fluence of 1.2 to 2.6 J/cm$^2$, respectively.

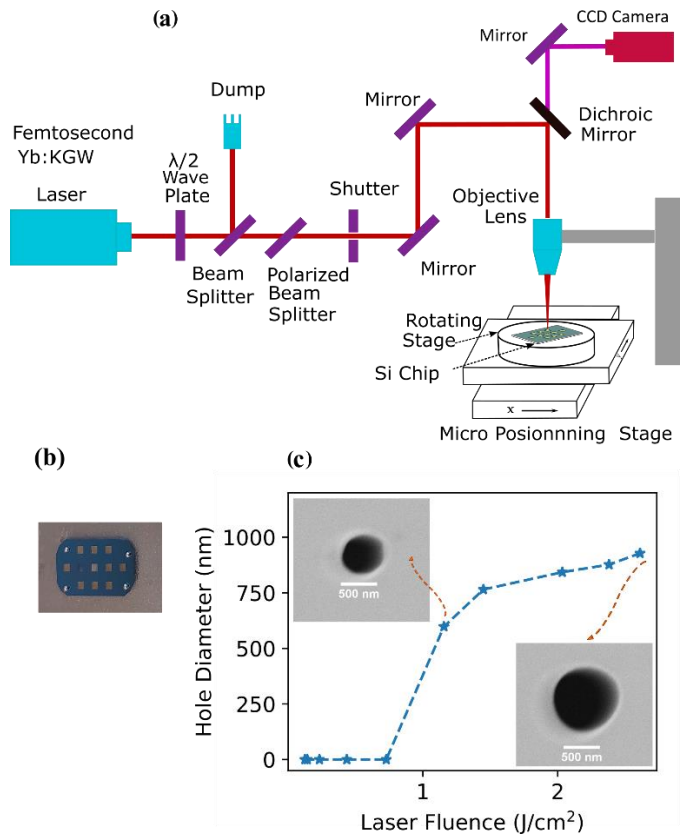

Fig. 1 (a) Schematic of femtosecond laser machining setup, (b) image of Si chip with 11 membranes with dimensions of 1.2 × 1.2 mm, (c) holes diameter drilled using single pulse laser as a function of laser fluence, inset show two SEM micrograph of drilled holes.

We observe that rastering continuously the laser over the membrane area to create etch lines is not a viable patterning solution. In this case stress concentrations eventually lead to cracking (Fig 2 a-c) when the etch lines extend to more than a few microns in length. For example, in Fig. 2 (a), a line is machined with a width of 1 µm, and a diagonal crack initiates after about 100 µm, at the circled area on the figure. Crack initiation is attributed to stress concentration since the heat input is limited (9.1 mW average power) and the since SiN is transparent at the machining laser wavelength (except for nonlinear absorption). Etching with larger beam diameters (10 µm instead of 1 µm) to limit the sharpness of the stress concentration point did not prevent cracking.

Destruction of the membrane during machining is avoided by carefully engineering and randomizing the laser etch pattern to avoid stress concentration. We find that individual single-pulse holes (right side of Fig 2 b, c) do not concentrate stress sufficiently for creating cracks, unless they are spaced closed enough to create a continuous line (left side of Fig 2 b, c). Following this logic, we also find that some larger geometries (such as the large circles in Fig 2 d) do not induce cracking if the etching pulses are applied randomly around the circle circumference, rather than as a continuous line. Consequently, we remove large geometries of SiN by first removing large circles (Fig 2 d, e). These circles are etched using 320 single-pulse holes distributed randomly on the circle, which has a 50 µm diameter in this case. This method of randomized pulse distribution was 100% successful in cutting circles of diameters as large as 400 µm without cracking. To cut nano beams on SiN membranes, an array of such circles was first cut, as can be seen in Fig. 2 g. Afterward, every four circles were connected using single-pulse holes distributed randomly as along the magenta dotted lines (marked with ii in Fig 2) to form round squares. Finally, the rounded squares were connected by machining the blue dotted lines using single-pulse non-randomized holes (Fig 2 e, marked with iii). Fig. 2h shows the final result of a straight 100 µm wide beam cut on a SiN membrane with this method.

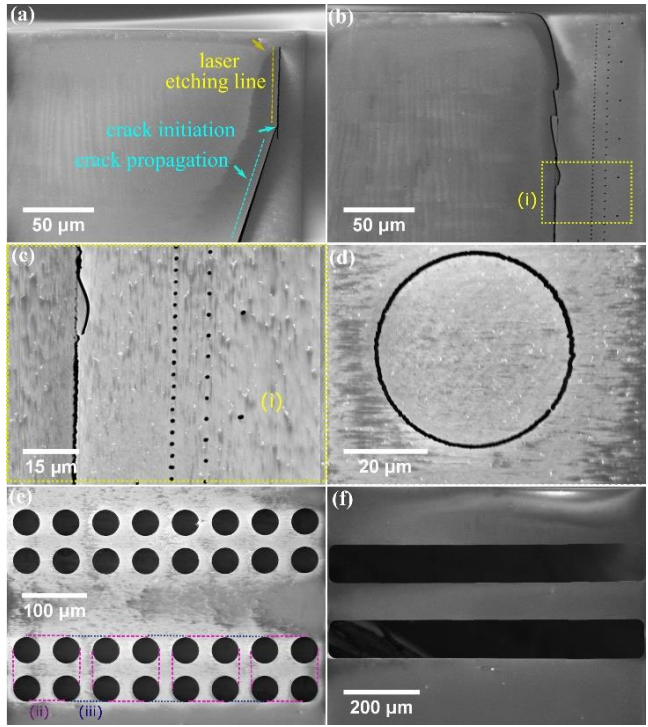

Fig. 2 Methodology of ultrafast laser micromachining of SiN membranes. (a) Crack initiation and fracture of SiN membrane after about 100 μm micromachining of a continuous line of 1 μm width, (b) array of holes drilled with single pulses with different distances. Drilled holes distance of 1 μm results in a cut-through slit, although cracks were initiated in a few spots and caused fracture, (c) selected area of (b) in higher magnification, (d) a circle cut successfully using 360 holes randomly distributed pulses (e) To form larger geometries, arrays of circles are first cut. Afterwards, every four circles are connected to form a round square (dashed line marked by (ii)). Finally, blue dotted lines are machined to form a large rectangular shape (marked with (iii)), and (f) a 100 μm wide straight beam was achieved by cutting the two rectangular shapes from the previous panel.

Using the described procedure, nine nanobeams (labelled i – ix in the remainder of this work) of various geometries are fabricated and are presented in Figs. 3. Beams i and ii (Fig. 3 a, b) extend only partially over the entire membrane area. Partial beams were first investigated to minimize cracking at the membrane clamping points, before the randomized pattern technique described above (Fig 2) was fully optimized. The other beams extend over the entire membrane length and are either tapered (iii – v) or straight (vi -ix). The tapered beams were cut with 15 and 50 µm widths, while four straight beams with widths of 100, 50, 15 and 7 µm were also machined. The higher magnification SEM micrograph of the selected area of Fig. 3 (f) and (g), shows that cuts are smooth and straight lines. One issue that remains is that some of the cut pieces get stuck due to electrostatic forces on the 100 and 50 µm wide straight beams. This can likely be resolved in the future by including a sacrificial layer on the SiN, by machining under vacuum to increase the travel length of debris, or by better post-machining cleaning procedure. In the current work, the fabricated beams are cleaned only by gentle blowing of dry nitrogen prior to subsequent mechanical characterization.

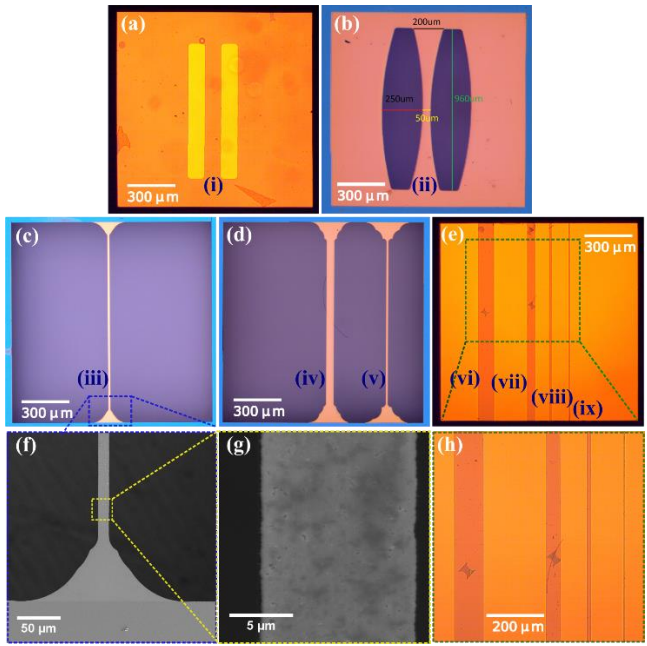

Fig 3: Pictures of all fabricated beams using ultrafast laser machining. (a) Partial beam, (b) curved machined beam, (c) 15 µm tapered beam, (d) 50 µm and 15 µm wide tapered beam (e) straight beams with 100 µm, 50 µm, 15 µm, and 7 µm width, respectively, (f-g) selected areas of (c) in higher magnifications, and (h) selected area of (e) in a higher magnification.

After fabrication, we characterize the vibration eigenmodes of all the nanobeams of Fig. 3 (except beams ii and iii that were damaged during handling) to determine how laser ablation affects mechanical dissipation, i.e., the mechanical quality factor. Characterization is performed in a custom-made high vacuum testing chamber ($\sim 7.5 \times 10^{-7}$ Torr). The four corners of the resonator's silicon frame are mounted on a steel substrate using eight 1 mm diameter magnets balls shown in Fig. 4 (b). Also attached to the back of the steel substrate is a shear piezo ceramic used to actuate the membrane. A fibre-optic interferometer (Fig. 4 a) records the resonator vibration amplitude and eigen frequency. The fibre optic is aligned on each machined beams using a digital microscope (Dino-Lite) imaging through a viewport at the top of the vacuum chamber, producing for example the image in Fig. 4 (b). The interferometer consists of a 1550-nm Orion™ laser, a 90:10 single-mode (approximately 10-µm mode field diameter) optical fibre coupler, and an amplified photo-detector (Thorlabs Inc. PDA20CS2). The laser output power (10.7 mW) is attenuated by the coupler (90%) and by a 5-dB in-line attenuator, such that the power incident on the sample is less than 35 µW. Signal from the photo-detector is sampled using a Zurich Instrument Ltd. MFLI lock-in amplifier (LIA).

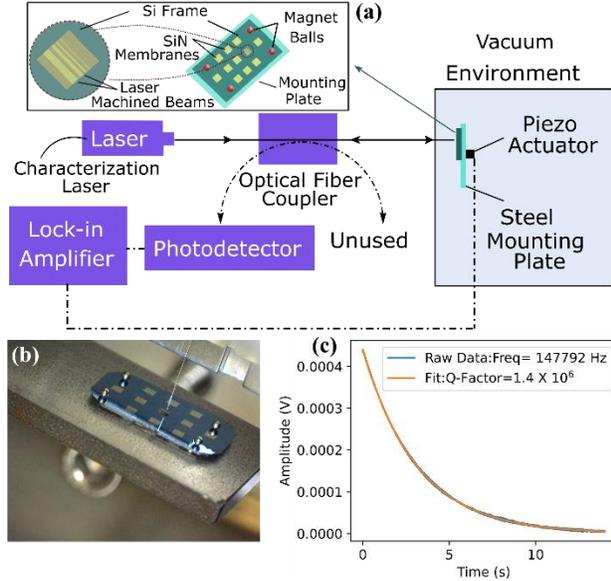

Fig. 4 Overview of the vibrational test setup. (a) Schematic of interferometer, resonator mounting, (b) picture of a Si chip mounted on a steel substrate in the high vacuum chamber with optical fibre readout (c) an example of vibration decay and quality factor measurement of a plain membrane with fitting a curve.

We first characterize the material dissipation of a plain, unprocessed, SiN membrane to provide a baseline for comparison with the laser-etched beams. Q factors for many modes of a plain membrane are presented in Fig. 5 (a), together with a damping dilution model[3] that predicts the membrane quality factor ($Q_{membrane}$) as a function of the material Q-factor ($Q_{mat}$) and a damping dilution factor ($\alpha_{dd,2D}$) :

$$Q_{membrane} = \alpha_{dd,2D} \times Q_{mat}, \qquad \text{eq. 1}$$

with

$$\alpha_{dd,2D} \approx \left[ \frac{\pi^2(n^2+j^2)}{12} \frac{E}{\sigma_{2D}} \left(\frac{h}{L}\right)^2 + \frac{1}{\sqrt{3}} \sqrt{\frac{E}{\sigma_{2D}}} \left(\frac{h}{L}\right) \right]^{-1}, \qquad \text{eq. 2}$$

and where $n$ and $j$ are the eigenmode orders, $E = 250$ GPa is elastic modulus of SiN, and $h = 90$ nm and $L = 1.2$ mm are the thickness and side length of the membrane. The built-in tensile stress $\sigma_{2d} = 70$ MPa is inferred from the measured eigenmode frequencies ($f_{m,n}$) using

$$f_{m,n} = \frac{\sqrt{m^2 + n^2}}{2L} \sqrt{\frac{\sigma_{2d}}{\rho}}, \qquad \text{eq. 3}$$

where $\sigma$ and $\rho$ are tensile stress and density. As is typical of SiN resonators dominated by clamping losses[6] the Q-factor of low order modes depends weakly on frequency. The material quality factor was extracted by fitting eq. 1 to the experimental data yielding $Q_{mat} \approx 2700$. Unsurprisingly, this is slightly lower than reported values for stoichiometric $Si_3N_4$ ($\approx 5000$)[6], thus confirming that our plain material behaves comparably to typical low-stress SiN.

We then analyse the eigenfrequencies of etched nanobeams to confirm that laser interaction does not lead to material modifications that affects the built-in tensile stress. The eigenfrequencies of one-dimensional beams is given by:

$$f_n = \frac{n}{2L} \sqrt{\frac{\sigma_{1d}}{\rho}}. \qquad \text{eq. 4}$$

If laser processing does not affect material stress, we expect stress relaxation to be caused only by the change from bi-axial to uniaxial stress, i.e., we expect $\sigma_{1d} = \sigma_{2d}(1 - \nu)$, where $\nu = 0.27$ is the Poisson ratio[22]. In this case, by inspection of eq. 3-4, we can expect the fundamental eigenfrequency ($f_1$) of beams or length $L$ to be equal to the fundamental eigenfrequency of membranes ($f_{1,1}$), multiplied by the following correction factor:

$$f_1 = f_{1,1} \times \sqrt{\frac{1 - \nu}{2}}. \qquad eq. 5$$

This is indeed what we observe in Fig. 5 (b) where all beams, regardless of their width, match closely the corrected membrane frequency, thus confirming that the differences in the frequency are related to geometry only. Note that in Fig. 5 (b) a length correction factor ($L_{partial}/L$) is also applied to partial and tapered beams. In the case of tapered beam $L_{partial}$ is equal to the non-tapered section of the beam.

While laser processing does not significantly affect the built-in tensile stress, we find that it has a a noticeable effect on material damping. This is illustrated by Fig. 5 (c), which shows that wider beams, in general, have higher maximum recorded Q-factors than narrower beams. This is contrary to the trend observed in SiN nanobeams fabricated by traditional lithography[23], which tend to show higher Q-factor for narrower beams. In our case, it is possible that laser processing creates a region of modified chemical composition of SiN near the edges, within which material damping is higher. Such regions would occupy a larger fraction of the total beam volume when the beams are narrower. Another possibility is that redeposition of material dust during laser etching has a lesser impact in wide beams, for which the center is further apart form the etching region and may receive less contaminants. Finally, we note that redeposition of large visible chunks of materials contaminants in beams (vi) and (vii) have a more dramatic effect on the measured Q-factors. These beams are therefore labelled as "contaminated beams" in Fig 5.

The measured Q factors of all the beams as a function of their frequency (Fig. 5 d) shows that material damping is more frequency dependant in etched beams than in plain membranes. This may indicate that damping goes from being clamping point-dominated in plain membranes (right term in Eq. 2), to being mode bending-dominated (left term in Eq. 6) in our etched nanobeams.

The difference in frequency dependence of plain membrane vs nanobeams is surprising at first since the damping dilution factor is very similar in 1D beam than in plain membranes, i.e.:

$$\alpha_{dd,1D} \approx \left[\frac{(n\pi)^2}{12}\frac{E}{\sigma}\left(\frac{h}{L}\right)^2 + \frac{1}{\sqrt{3}}\sqrt{\frac{E}{\sigma}}\left(\frac{h}{L}\right)\right]^{-1}. \qquad \text{eq. 6}$$

In eq. 6, the first (frequency dependant) term accounts for material damping from the general sinusoidal mode shape, while the second (frequency independent) term, which usually dominates for low *n* values, accounts for damping from sharp bending at the clamping point. The frequency dependence of damping observed in Fig 5 (d) indicates that our process likely makes material damping at the center more important than at the clamping point. This is also supported by the fact that 15 μm wide beams that are either tapered (v) or straight (viii) have very similar response in Fig 5 (d), although recent study show that tapered beams should show have lower Q-factors [24] if clamping point damping dominates. It is possible that material re-deposition or heat loading is worst at the center of the membrane than at the clamping points, which would explain the reduced relative impact of clamping point damping. Another possibility is that catastrophic material redeposition occurs randomly over the beams during laser etching. In this case, material redeposition is more probable somewhere over the length of the beams than exactly at the clamping points.

Nevertheless, the fact that all beams in Fig 5 (d) reach Q-factors on the order of 0.1 – 1 million indicates that our etch method is promising, as it does not catastrophically affect the material properties in the most sensitive region—i.e., the clamping point. Better cleaning procedures, or ablation under vacuum, will likely allow the material Q to stay high over the rest of the beam area, although this will require further test to confirm experimentally.

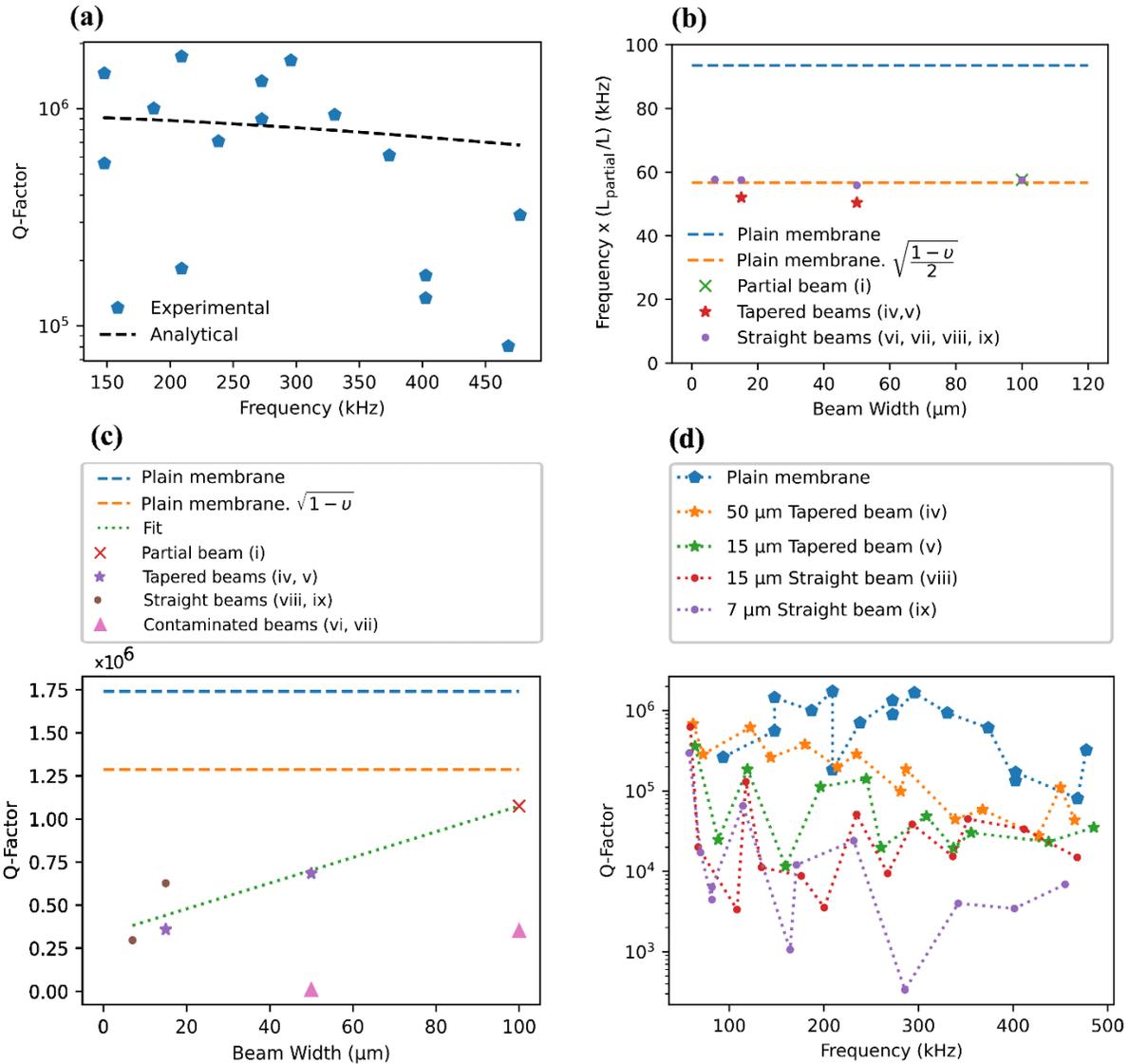

Fig 5: Characterization of resonance frequency and quality factor of micromachined beams. (a) the material damping $Q_{mat}$ was extracted by the fitting the analytically predicted frequencies and Q factors to the experimentally measured values of the plain membrane, the $Q_{mat}$ was identified to be 2710, (b) normalized fundamental mode of beams as a function of their width, the frequency of the fundamental mode of the plain membrane was presented as a guide with dotted lines, (c) the highest Q factors of the beams as a function of the width of the beam, the highest

measured Q factor of the plain membrane was given as a guide, and (d) Q factor of all laser-etched beams as a function of frequency.

## Author Declarations

The authors have no conflicts of interest to declare.

## Data Availability

The data that support the findings of this study are available from the corresponding author upon reasonable request.

20. Snell N, Zhang C, Mu G, Bouchard A, St-Gelais R. Heat Transport in Silicon Nitride Drum Resonators and its Influence on Thermal Fluctuation-Induced Frequency Noise. *Physical Review Applied*. 4AD;17(4):044019. doi:10.1103/PhysRevApplied.17.044019

21. Mu, G., Snell, N., Zhang, C., Xie, X., Tahvildari, R., Weck, A., Godin, M. and St-Gelais, R Observation of Silicon Nitride Nanomechanical Resonator Actuation Using Capacitive Substrate Excitation. *arXiv preprint arXiv:211209303*. Published online 2021.

22. St-Gelais R, Bernard S, Reinhardt C, Sankey JC. Swept-Frequency Drumhead Optomechanical Resonators. *ACS Photonics*. 2019;6(2):525-530. doi:10.1021/acsphotonics.8b01519

23. Schmid S, Jensen KD, Nielsen KH, Boisen A. Damping mechanisms in high-Q micro and nanomechanical string resonators. *Physical Review B*. 2011;84(16):165307.

24. Bereyhi MohammadJ, Beccari A, Fedorov SA, et al. Clamp-Tapering Increases the Quality Factor of Stressed Nanobeams. *Nano Lett*. 2019;19(4):2329-2333. doi:10.1021/acs.nanolett.8b04942

**17 |** P a g e